\documentclass[twocolumn, showpacs, prb]{revtex4}
\usepackage{amsmath,amssymb}
\usepackage{graphicx}
\usepackage{epstopdf}
\usepackage{bm}
\usepackage[usenames,dvipsnames]{xcolor}

\begin{document}


\title{Ab-initio prediction of magnetoelectricity \\ 
in infinite-layer CaFeO$_{2}$ and MgFeO$_{2}$ } 


\author{Kunihiko Yamauchi$^{1}$}
\author{Tamio Oguchi$^{1,2}$}
\author{Silvia Picozzi$^3$}%
 


\affiliation{
1. ISIR-SANKEN, 
 Osaka University, 8-1 Mihogaoka, Ibaraki, Osaka, 567-0047, Japan \\
\color{black} 2. CREST, JST,  4-1-8 Honcho, Kawaguchi, Saitama, 332-0012, Japan \color{black} \\
3. Consiglio Nazionale delle Ricerche (CNR-SPIN), 67100 L'Aquila, Italy 
}


\date{\today}
\newcommand{\ca}{CaFeO$_{2}$}
\newcommand{\ma}{MgFeO$_{2}$}
\newcommand{\ba}{Ba$_{2}$CoGe$_{2}$O$_{7}$}
\newcommand{\bcgo}{Ba$_{2}$CoGe$_{2}$O$_{7}$}
\newcommand{\beq}{\begin{eqnarray}}
\newcommand{\eeq}{\end{eqnarray}}
\newcommand{\degree}{\ensuremath{^\circ}}

\begin{abstract}


Density functional based simulations are employed to explore  magnetoelectric effects in iron-based oxides, showing  a unique layered structure. 
We theoretically predict  \ca\   to be  a promising magnetoelectric, showing magnetically-controlled large electric polarization, 
possibly  even above room temperature. 
The cross coupling between magnetic and dipolar degrees of freedom needs, as main ingredients,  Fe-site spin-orbit coupling and a spin-dependent O $p$ - Fe $d$ hybridization, along with structural constraints 
 related to the non-centrosymmetric point group and the peculiar geometry characterized by ``flattened" FeO$_{4}$ tetrahedrons. 
In order to enhance  magnetoelectric effects, we performed a materials-design leading to a novel and optimized system, \ma, where the 
larger O$_{4}$ tetrahedral distortion leads to a stronger polarization. 
 
\end{abstract}

\pacs{Valid PACS appear here}
\maketitle

\section{Introduction}
\color{black}
Multiferroic oxides represent a playground for magnetoelectric (ME) effect to arise, 
offering different ways in which magnetization ($M$) and the  ferroelectric polarization ($P$) couple. 
\color{black}
%
Microscopically, ME mechanisms can be classified into 
either the relativistic ``inverse Dzyaloshinskii-Moriya (DM) mechanism'',\cite{sergienko.prb2006}
\color{black}
often related to  spin-current mechanism \cite{KNB}, 
 \color{black}
showing $\bm{P}\propto\sum_{ij}\bm{e}_{ij}\times(\bm{S}_{i}\times\bm{S}_{j})$ between neighboring spins
connected by a vector $\bm{e}_{ij}$, 
or
non-relativistic  inverse Goodenough-Kanamori (or exchange-striction) mechanism \cite{picozzi.homno3} showing
$\bm{P}\propto\sum_{ij}J_{ij}(\bm{S}_{i}\cdot\bm{S}_{j})$ with exchange integral $J_{ij}$. 
\color{black}
In many multiferroics, 
\color{black}
such as MnWO$_{4}$ and \color{black} 
CaMn$_{7}$O$_{12}$, 
\color{black}
even the latter ME mechanism is indirectly linked to 
\color{black}
relativistic effects, as the spin direction is determined by DM interaction and magnetic anisotropy, 
\color{black}
which, in turn, cause the ME effect.\cite{whangbo2012.camno7, solovyev2013.mnwo4} 
\color{black}
%
 Recently, an alternative mechanism, denoted as {\it ``spin-dependent $p$-$d$ hybridization''}  
 was proposed.\cite{arima.jpsj2007, jia.nagaosa.prb2007} 
There, spin-orbit-coupling (SOC)
affects the $p$-$d$ hybridization between the transition metal (TM) and the surrounding anions, inducing an electric
polarization  $\bm{P}\propto\sum_{ij}(\bm{S}_{i}\,\cdot\,\bm e_{j}')^{2}\,\bm e_{j}'$, where $\bm e_{j}'$ labels the
vectors connecting the TM to the ligand ions. 
\color{black}
In addition to explaining 
 \color{black} 
 ferroelectricity in CuFeO$_{2}$,\cite{arima.jpsj2007}  
it has been more recently reported that the mechanism can be responsible for the polarization observed
in \ba\ (BCGO), where two neighboring Co spins are aligned in an antiferromagnetic (AFM) configuration,\cite{tokura.prl2010}  
as confirmed by a previous DFT study \color{black} by some of us.\cite{yamauchi.bcgo} 
\color{black}
There, in the CoO$_{4}$ tetrahedral coordination, the occupied {\it non-bonding} $x^{2}-y^{2}$ state for minority spins is mixed with unoccupied {\it bonding} $yz$ and $zx$ states through SOC,  leading to {\it asymmetric pd hybridization} and, in turn, to a net polarization.  
%
\color{black}
We also remark that 
\color{black}
a relevant ME mechanism \color{black}
was also found \color{black}
in magnetite, Fe$_{3}$O$_{4}$, 
where 
the complicated charge-ordering pattern with  polar $Cc$ space group induces a large spontaneous polarization, 
whereas SOC causes a small  spin-dependent change in the polarization under the ferrimagnetic order.\cite{yamauchi.magmagmag}  

By means of a  theoretical analysis performed for BCGO and Fe$_{3}$O$_{4}$, 
\color{black}
and aiming at the discovery of \color{black}
novel ME materials, we identified 
the requirements for the spin-dependent $pd$ hybridization mechanism:  
{i)} a non-polar, but non-centrosymmetric, point symmetry at magnetic sites, whose symmetry is lowered into a polar group under a specified magnetic order;
{ii)} weak magnetic anisotropy (so that spins easily follow the magnetic field), at least in a plane, 
{iii)} strong $pd$ hybridization and strong SOC needed to achieve a large ME effect. 
Whereas the third condition can be fulfilled by using either 4$d$ or 5$d$ elements instead of 3$d$ transition metals, 
our approach here consists in  exploring different crystal structures to enhance the asymmetric $pd$ hybridization to realize larger polarization. 

%
\begin{figure}[ht]
\vspace{-0.2cm}
{
\includegraphics[width=1.00\columnwidth, angle=0]{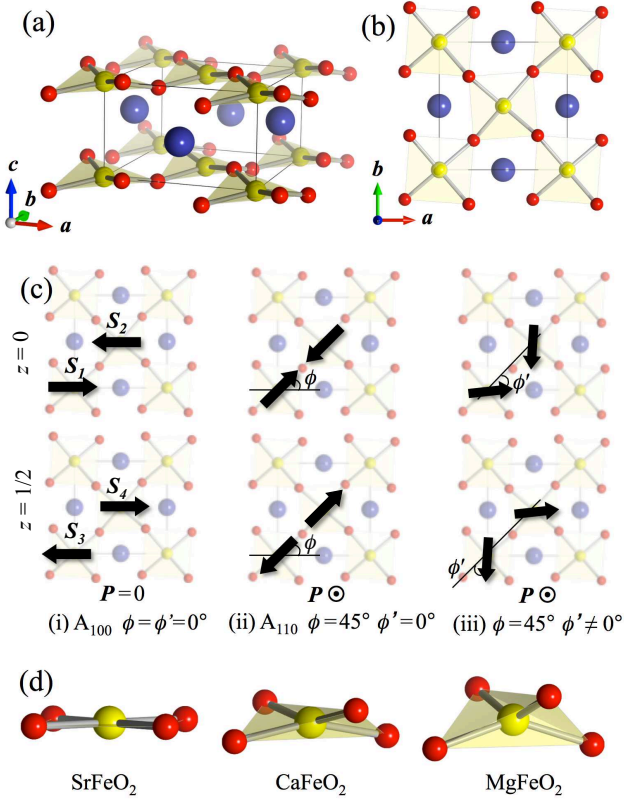}
}
\caption{\label{fig:crys} 
(a) Crystal structure and (b) projected image in the $ab$ plane of CaFeO$_{2}$ with the $P$-$42_{1}m$ space group; 
 Fe ions (located in flat O$_{4}$ tetrahedrons) lie in $c$=0 planes whereas
Ca ion lies in $c$=1/2 planes. 
(c) Fe spin configurations: (i), (ii) collinear AFM and (iii) non-collinear spin-canted under
applied $\bm H$//1-10. Note that the AFM inter-layer coupling is not shown here. 
(d) 
\color{black}Increase of buckling and \color{black}
development of MO$_{4}$ tetrahedron with respect to the planar square coordination by substituting Sr  by Ca and Mg. 
}
\vspace{-0.3cm}
\end{figure}

Our strategy is rather simple. We searched for oxides  having a   space group similar to BCGO 
and identified  
CaFeO$_{2}$ as a promising candidate.  
Indeed, \ca\  crystallizes in {\em non-centrosymmetric and non-polar} $P\overline{4}2_{1}m$ (\#113) tetragonal layered structure. 
The crystal structure shows peculiar FeO$_{4}$ squares heavily distorted towards tetrahedrons (Fig. \ref{fig:crys} (a)). 
Experimentally CaFeO$_{2}$ can be synthesized 
\color{black}
by starting from the precursor Brownmillerite CaFeO$_{2.5}$ and removing only apical oxygens in FeO$_{6}$ octahedrons, by means of  reductive hydride CaH$_{2}$. 
\color{black}
%
\color{black}
The recently synthesized \ca\ uniquely shows  corrugation in FeO$_{2}$ planes, 
with the O ion  slightly deviated (by 0.32\AA) from the $ab$ plane.\cite{tassel}
\color{black}
Such distortion has been neither reported in the infinite-layer cuprate ACuO$_{2}$ (A= Ca, Sr, Ba), 
well studied in the context of high-temperature superconductivity\cite{azuma.nature}, 
nor in  SrFeO$_{2}$, having high-spin Fe ions in planar square coordination.\cite{tsujimoto} 

We note that CaFeO$_{2}$ exhibits G-type antiferromagnetic order (\color{black}showing \color{black}  both in-plane and inter-plane AFM coupling)  
below a remarkably high N\'eel temperature, $T_{\rm N}$=420K. 
Four Fe atoms in the unit cell, Fe1 at (0, 0, 0), Fe2 at (1/2, 1/2, 0), Fe3 at (0, 0, 1/2), and Fe4 at (1/2, 1/2, 1/2), 
are responsible for the magnetism (the magnetic unit cell is doubled along the $c$ axis 
compared to the structural unit cell). 

In this study, by means of DFT calculations, we first investigate the steric
 effect \color{black}
 induced by the \color{black} 
  A site substitution (A= Mg, Ca, Sr, Ba) in AFeO$_{2}$ and 
the stability of the infinite-layer structure; then,  
we investigate  ME effects caused by varying the magnetic order under SOC.  
\color{black} 
As BCGO can be considered a prototypical case of spin-dependent $p$-$d$ hybridization inducing ME effects, 
we'll compare -- whenever possible -- our findings for CaFeO$_{2}$-based systems with our reference material, BCGO. 
\color{black} 


\begin{figure}[!h]
\includegraphics[width=0.99\columnwidth]{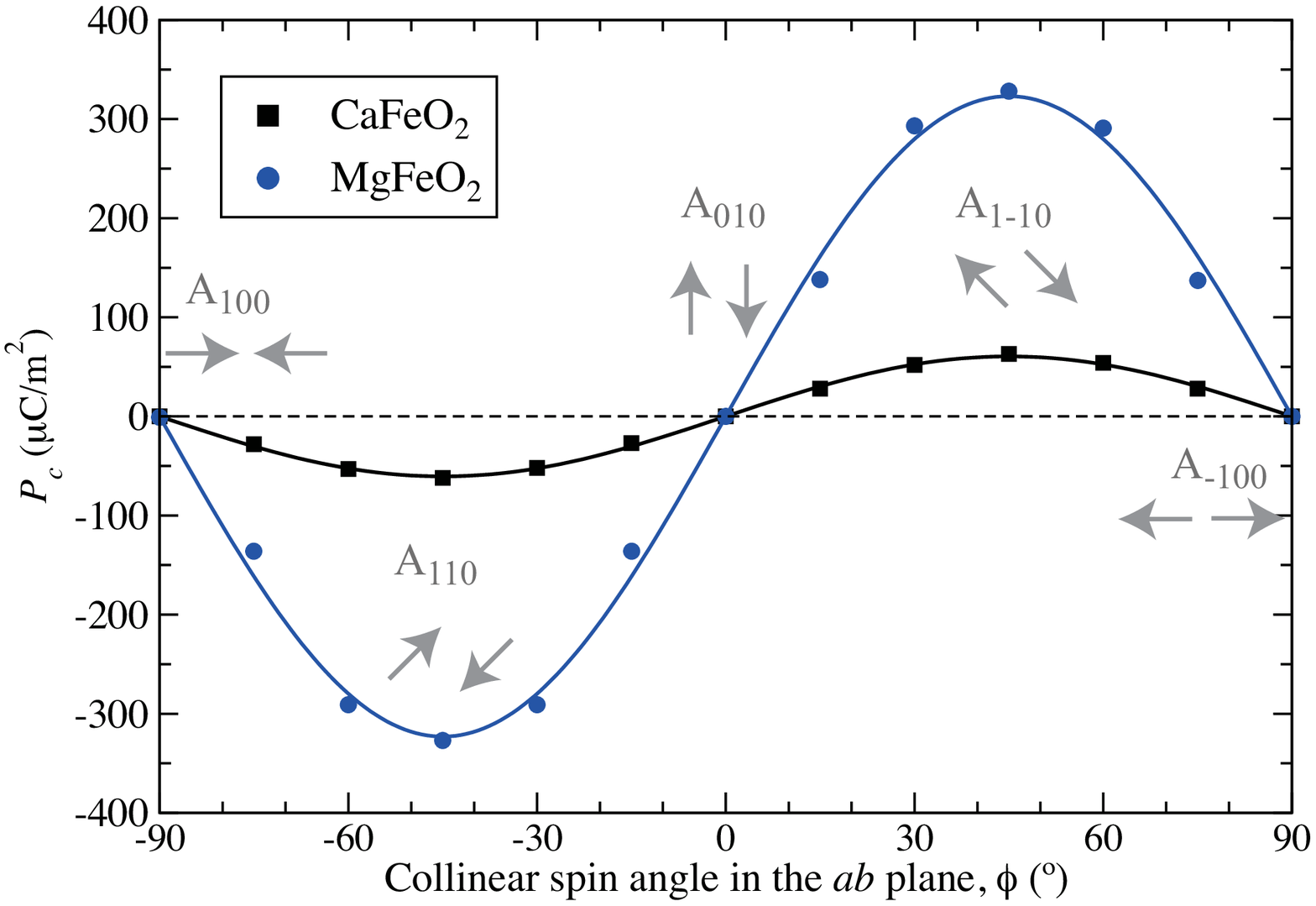}
\\
\vspace{0.2cm}
\includegraphics[width=0.99\columnwidth]{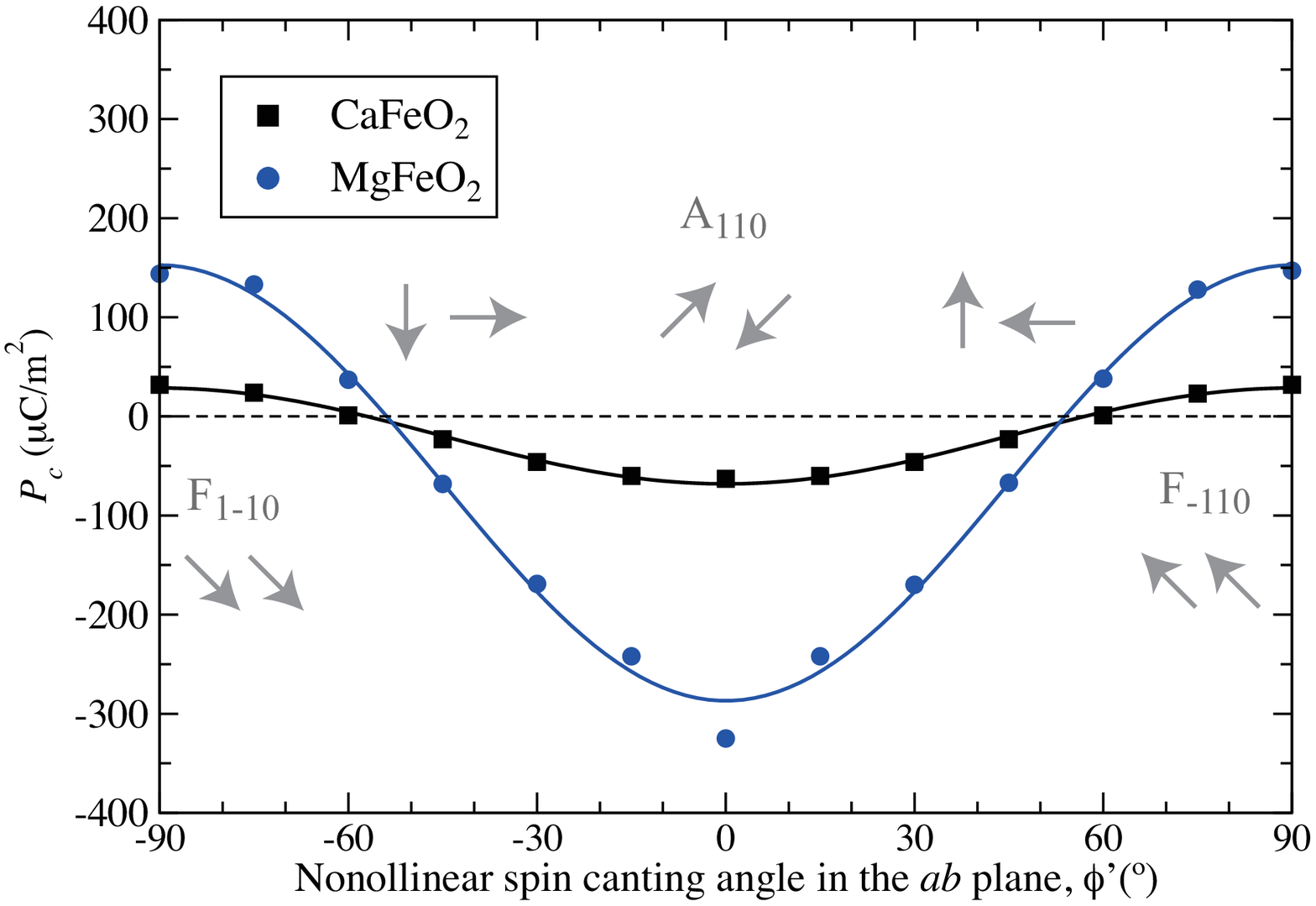}
\caption{\label{fig:DFT} 
(a) \color{black} First-principles $P_{c}$ evaluated \color{black} 
 as a function of the 
 spin angle $\phi$ in the $ab$ plane,  fitted to  sine-like curves. 
The spin configurations in the $ab$ plane is \color{black} schematically \color{black} shown by grey arrows. 
(b) 
$\bm P_{c}$  as a function of the 
spin-canting angle $\phi'$ in the $ab$ plane,  \color{black} (non-collinear configuration) \color{black} 
fitted to cosine-like curves (solid line). 
}
\end{figure}
%

\section{Symmetry Analysis}
\color{black}
We start by performing a symmetry study based on group theory and then consider its consequences on the analysis in terms of the Landau theory of phase transitions.\cite{Landau}  
 \color{black}
 We note that, although the space group of crystal structure of CaFeO$_{2}$ is the same as  BCGO, 
the G-type AFM order may give a different ME behavior compared to the C-type (only in-plane AFM coupling) AFM order in BCGO. 
\color{black}
In the parent $P\overline{4}2_{1}m1'$  space group with eight symmetry operations
\{$E$, $C_{2(z)}$, 2$S_{4}$, 2$C_{2(x,y)}$, 2$\sigma_{d}$\}
plus time-reversal \{1'\}, 
the occurrence of a spin transition lowers the symmetry. 
We define the order parameters  
${\bm F}={\bm S_{1}}+{\bm S_{2}}+{\bm S_{3}}+{\bm S_{4}}$ and  ${\bm A}={\bm S_{1}}-{\bm S_{2}}-{\bm S_{3}}+{\bm S_{4}}$ as the ferromagnetic (FM) and G-type AFM
combination of Fe${1..4}$ spins, respectively. 
Since the G-type AFM order doubles the unit cell along the $c$ direction, we have to take into account the $c$ translation \{$c=$+(001)\} in this analysis ($i.e.$ AFM order is not invariant under the $c$ translation).  
\begin{table}[tdp]
\vspace{-0.2cm}
\caption{
Matrices of the generators of space group  $P\overline{4}2_{1}m1'$  in the representations spanned by 
\color{black} 
order parameters: 
\color{black} 
$F$, $A$, and $P$. 
The group elements denote the identity, $\pi$-rotation, $\pi/2$-rotoinversion $S_{4}^{-}$(=$IC_{4z}^{-}$), screw $C_{2y}$+$(\frac{1}{2}
\frac{1}{2} 0)$, time-reversal $1'$ and translation $c$. 
\color{black}
The notation for the irreducible representation (IR) follows from the ISODISTORT software.\cite{isodistort}  
\color{black}
Note that the G-type AFM order ($A_{a}$, $A_{b}$, $A_{c}$) here shows different symmetry properties with respect to the C-AFM order in BCGO.\cite{yamauchi.bcgo}
\color{black}
\label{table:group}}
\vspace{-0.3cm}
\begin{center}
{\setlength{\tabcolsep}{2pt}\footnotesize
\begin{tabular}{|c|cccccc|c|}
\hline
	& 	$E$  &$C_{2z}$ & $S_{4}^{-}$ & $C_{2y}$&$1'$ & $c$ & IR  \\
     \hline
%
$\begin{array}{c}F_{a} \\F_{b}\end{array}$ &
$\begin{array}{c} 1 \\ 1 \end{array}$ &
$\begin{array}{c} -1 \\ -1 \end{array}$ &
$\begin{bmatrix}0 & 1 \\-1 & 0\end{bmatrix}$ &
$\begin{array}{c} -1 \\ 1 \end{array}$ &
$\begin{array}{c} -1 \\ -1 \end{array}$ &
$\begin{array}{c} 1 \\ 1 \end{array}$ & 
$\begin{array}{c}m\Gamma_{5}E_{1}^{*}a \\ m\Gamma_{5}E_{1}^{*}b\end{array}$ \\
$F_{c}$&
1&1&1&$-$1&$-$1	& 1 &
$m\Gamma_{4}A$ \\
\hline
%
$\begin{array}{c}A_{a} \\A_{b}\end{array}$ &
$\begin{array}{c} 1 \\ 1 \end{array}$ &
$\begin{array}{c} -1 \\ -1 \end{array}$ &
$\begin{bmatrix}0 & -1 \\1 & 0\end{bmatrix}$ &
$\begin{array}{c} -1 \\ 1 \end{array}$ &
$\begin{array}{c} -1 \\ -1 \end{array}$ &
$\begin{array}{c} -1 \\ -1 \end{array}$ & 
$\begin{array}{c}mZ_{5}E^{*}_{(1,2)} \\ mZ_{5}E^{*}_{(1,2)}\end{array}$ \\
$A_{c}$&1&1&$-$1&$-$1&1&$-$1&$mZ_{1}Aa$	\\
\hline

%
$\begin{array}{c}P_{a} \\P_{b}\end{array}$ &
$\begin{array}{c} 1 \\ 1 \end{array}$ &
$\begin{array}{c} -1 \\ -1 \end{array}$ &
$\begin{bmatrix}0 & -1 \\1 & 0\end{bmatrix}$ &
$\begin{array}{c} -1 \\ 1 \end{array}$ &
$\begin{array}{c} 1 \\ 1 \end{array}$ &
$\begin{array}{c} 1 \\ 1 \end{array}$ & 
$\begin{array}{c}\Gamma_{5} \\ \Gamma_{5} \end{array}$ \\
$P_{c}$&1&1&$-$1&$-$1&1&1	&$\Gamma_{3}$\\
\hline
\end{tabular}}
\end{center}
\label{default}
\vspace{-0.7cm}
\end{table}

Using the transformation rules reported in Table \ref{table:group} 
\color{black}
and considering 
\color{black} 
all the possible ME 
coupling terms of the form ${\bm P} \cdot {\bm M}^{2}$, which are invariant under any symmetry operation,  
\color{black}
one can derive \color{black} 
the  simple expression 
\color{black} 
for the thermodynamic free energy: 
\color{black} 
%
%
%
\begin{eqnarray}
 &F_{\rm ME} = 
    c_{\rm A}P_{c}A_{a}A_{b} + c_{\rm F}P_{c}F_{a}F_{b}. 
\end{eqnarray}
The dielectric energy is written as $F_{\rm DE} =- {\bm P}^{2}/2\chi $. 
Here,  $c_A, c_{F}$ and $\chi$ (from now on, set as 1) are constants.
%
When evaluating $\bm P$ at the minima of $F=F_{\rm ME}+F_{\rm DE}$, one gets
%
%
\begin{align}\label{eq:P}
P_{a}=P_{b}=0,  \qquad 
P_{c}=c_{\rm A}A_{a}A_{b} + c_{\rm F}F_{a}F_{b}.  
\end{align}


We observe that this result means that $P_{c}$ can originate from either FM or AFM order, but not from their combination.\cite{footnote1}   

\color{black}
Let's now focus on $P_{c}$ as induced by an applied magnetic field causing a canted AFM spin arrangement. 
In this situation, as a first step, we rotate counterclock-wise
\color{black}
 four Fe spins  in the $ab$ plane, keeping the AFM order (with an angle $\phi$ from the $a$ axis). 
 \color{black}
 We later rotate \color{black} the spins by an angle $\phi'$ (see Fig. \ref{fig:crys}). 
\color{black} Under these conditions, it is possible to write: \color{black}
${\bm S_{1}}={\bm S_{4}}=S\,(\,\cos(\phi+\phi'),\, \sin(\,\phi+\phi'),\, 0)$ and 
${\bm S_{2}}={\bm S_{3}}=S(\,-\cos(\phi-\phi'),\,-\sin(\phi-\phi'),\, 0)$, so that  
%
\begin{equation}
P_{c}= 4 a S^{2} \sin2\phi \left( \cos2\phi'+b\right), \label{eqs:Pangle}
\end{equation}
where $a=c_{\rm A}+c_{\rm F}$ and $b=c_{\rm A}+c_{\rm F} -2$. 
%


\color{black} 
On the basis of Eq. \ref{eqs:Pangle}, we observe that a spontaneous $P_{c}$ arises in the presence of \color{black}
the A$_{110}$ or A$_{1-10}$ ($\phi$=$\pm$45$^{\circ}$) orders, but not of the A$_{100}$ (A$_{010}$)
one. 
In analogy with  previous studies \cite{yamauchi.bcgo, yamauchi.magmagmag}, the non-magnetic group lacks  inversion symmetry, but the
symmetries which prohibit $P_{c}$ (e.g. $C_{2y}$ rotation) are broken by the A$_{110}$ magnetic order. 
Finally, Eq. (\ref{eqs:Pangle}) gives 
\color{black}
a simple behavior as a function of  $\phi$ and $\phi'$, such as 
\color{black}
$P_{c}(\phi)\propto \sin2\phi$ and $P_{c}(\phi')\propto \cos2\phi'$. 
%
This result 
\color{black} 
is different \color{black} 
from the $P$ induced in C-AFM BCGO\cite{yamauchi.bcgo, tokura.prl2010}, where 
$P_{c}^{\rm BCGO}(\phi')\propto \cos(2\phi'-\beta)$, 
 the phase shift $\beta$ depending on the non-zero $c_{\rm AF}$ coefficient. 
Notably, the phase shift of the cosine curve, $\beta$, is lost in G-AFM CaFeO$_{2}$. 
Physically, a close inspection of the magnetic and ionic configuration shows that 
the cancellation is due to the C type interlayer AFM coupling and the consequent spin canting pattern in different  layers (see Fig. \ref{fig:crys}(c)-(iii)). 


\color{black}

\section{DFT analysis}
\color{black}
To support the symmetry analysis by a complimentary approach, 
\color{black} 
we performed DFT
calculations on $A$FeO$_{2}$ ($A$ = Mg, Ca, Sr, Ba)
by using the VASP code\cite{vasp} with GGA-PBE+$U$ potential ($U$=4 eV for  Fe-$d$ state, taken from Ref.\cite{tassel}).  
Note that bare GGA calculations result in quasi-metallic state with small gap, $e. g.$ $E_{g}\lesssim$0.1eV in MgFeO$_{2}$. 
\color{black}
For a more accurate treatment of 
\color{black}
the magnetic anisotropy and the ME effect, the SOC term was computed self-consistently inside each atomic sphere (see Ref. \cite{kunihiko.linipo4}). 
\color{black}
%
%
%
\begin{table}[h!]
\vspace{-0.5cm}
\caption{
\color{black}
Theoretically optimized unit-cell lattice parameters $a$ and $c$ (\AA), Fe-O bond length (\AA) and Fe-O-Fe bond angle  ($^{\circ}$) , along with the corresponding experimental values. 
\color{black}
\label{table:struc}
}
\begin{center}
\begin{tabular}{|cl|cc|cc|}
\hline
	& 	&$a$ & $c$ &$d_{\rm Fe-O}$ & $\angle_{\rm FeOFe}$\\
     \hline
SrFeO$_{2}$ 				&\footnotesize DFT						&3.994	&3.430	&1.997&180  \\
\footnotesize$P4/mmm$		&\footnotesize EXP\cite{tsujimoto.nature.srfeo2}&3.985	&3.458	&1.993&180 \\
\hline
CaFeO$_{2}$				&\footnotesize DFT			&5.518 & 3.328 & 1.993 & 156.27\\
\footnotesize$P$-$42_{1}m$	&\footnotesize EXP\cite{tassel}	&5.507 & 3.356 &1.978 & 159.69  \\
			\hline
MgFeO$_{2}$				&\footnotesize DFT			&5.278 & 3.020 & 1.999 & 137.93  \\
\hline
\end{tabular}
\end{center}
\vspace{-0.5cm}
\end{table}
%

\color{black}
The structural optimization of $A$FeO$_{2}$ was done as follows: starting with the experimental structure of CaFeO$_{2}$,\cite{tassel} internal atomic coordinates as well as lattice parameters were fully optimized under G-AFM configuration without SOC, until forces acting on atoms were less than 1$\times10^{-3}$ eV/\AA. 

We found  
MgFeO$_{2}$ and CaFeO$_{2}$ to show a corrugation in the FeO$_{2}$ layer in  non-centrosymmetric $P$-$42_{1}m$ structure, 
whereas  SrFeO$_{2}$ and BaFeO$_{2}$ show a ``flat'' layer in centrosymmetric $P4/mmm$ structure. 
The optimized structural parameters are shown in Tab.\ref{table:struc}, 
where the deviation of lattice parameters from the experimental value is less than 2\%. 
\color{black}
In the FeO$_{4}$ tetrahedra of MgFeO$_{2}$ and CaFeO$_{2}$, the JT-active Fe$^{2+}$ ($d^{6}$)  ion shows  
$e_{g}^{2\uparrow}$$t_{2g}^{3\uparrow}e_{g}^{1\downarrow}$$t_{2g}^{0\downarrow}$ occupied orbital states, 
where the almost-non-bonding $3z^{2}-r^{2}$ states perpendicular to the $ab$ plane is the lowest energy orbital state. 
(Here we choose a local frame $\bf x$//$\bf a$, $\bf y$//$\bf b$, $\bf z$//$\bf c$.) 

%
\begin{table}[h!]
\vspace{-0.5cm}
\caption{
\color{black}
Top: Total energy difference (meV/Fe) between several magnetic orders and inter-site magnetic coupling constants $J_{ij}$ obtained without SOC: 
nearest neighboring coupling $J_{ij}^{1 \rm \parallel}$  in layer,  $J_{ij}^{1 \rm \perp}$ inter layer, and 
second nearest coupling  $J_{ij}^{2\rm \parallel}$ in layer. 
The N\'eel Temperature of G-AFM order, $T_{\rm N}^{\rm G}$ (K), is obtained from  mean-field approximation 
$k_{\rm B} T_{\rm N}^{\rm G} \sim 2/3 (4 J_{ij}^{1 \rm \parallel} + 2 J_{ij}^{1 \rm \perp})$, 
where $k_{\rm B}$ is the Boltzmann constant. 
%
Bottom: Magnetic anisotropy energy (MAE) (meV/Fe) obtained by comparing the total energy with different spin directions under SOC
 in the GGA+$U$ scheme. Spin and orbital moment, $S$ and $L$ ($\mu_{\rm B}$), are also reported for $S$//(100).
\label{table:mae}
}
\begin{center}
\begin{tabular}{|c|cccc|ccc|c|}
\hline
	& 	FM & A-AFM & C-AFM & G-AFM & $J_{ij}^{1 \rm \parallel}$ &  $J_{ij}^{1 \rm \perp}$ &  $J_{ij}^{2\rm \parallel}$ & $T_{\rm N}^{\rm G}$\\
     \hline
CaFeO$_{2}$	&    0.0 &   -7.8 &  -83.2 &  -93.6 &  21.13 &   4.54 &  -0.16 &724\\
MgFeO$_{2}$	&    0.0 &  -10.5 &  -60.8 &  -64.3 &  14.32 &   3.52 &   0.44&498   \\
\hline
\end{tabular}
\end{center}
\begin{center}
\begin{tabular}{|c|ccc|cc|c|}
\hline
	& 	$E$(100) & $E$(110) & $E$(001) & $S$ & $L$ \\
     \hline
CaFeO$_{2}$	&0 & 0.00 & +2.03& 3.61 &0.09  \\
MgFeO$_{2}$	&0 & +0.60 & +2.90 & 3.61 &0.10  \\
\hline
\end{tabular}
\end{center}
%
\vspace{-0.5cm}
\end{table}
\color{black}
The magnetic stability is shown in Tab.\ref{table:mae} \color{black}
as evaluated by comparing total energies for \color{black} 
 ferromagnetic (FM) and AFM orders with A-, C-, and G-type configurations. 
For both CaFeO$_{2}$ and MgFeO$_{2}$, the most stable magnetic order  is G-type AFM order, 
consistent with the experimental observation in CaFeO$_{2}$.\cite{tassel}  
However, in MgFeO$_{2}$,  C-AFM and G-AFM are  rather close in energy, due to the frustration of $J_{ij}^{1 \rm \perp}$ and  $J_{ij}^{2\rm \parallel}$. 
In order to explain the antiferromagnetic superexchange behavior, one can exploit Goodenough-Kanamori rules in the case of  $d^{6}$-O-$d^{6}$  straight bond.\cite{goodenoughkanamori} 
While CaFeO$_{2}$ shows a stable G-AFM order with a high N\'eel Temperature: based on mean-field approximation, $T_N^{\rm calc}$ highly overestimates experimental $T_N^{\rm exp}$=420K. 
\color{black} On the other hand, 
MgFeO$_{2}$ shows slightly lower $T_N^{\rm calc}$ (Tab.\ref{table:mae}). 
The reduction of AFM stability is likely caused by the smaller Fe-O-Fe bond angle, which weakens the Goodenough-Kanamori super-exchange. 
\color{black}

The calculated small magnetic anisotropy in the easy $ab$ plane and hard $c$ axis (Tab. \ref{table:mae}) 
grants an  
easy control over the spins by applied magnetic field, $\bm H$; 
\color{black} even when $\bm H$ is small, we expect the spins to arrange perpendicularly to the applied filed, 
possibly adopting a small canting to reduce the Zeeman energy.  \color{black}

Assuming a collinear AFM spin arrangement, we simultaneously rotate the Fe spins in the $ab$ plane. 
The ME effect \color{black} is then evaluated \color{black} as the change of $\bm P$ (calculated by Berry phase approach\cite{berry})
induced by the rotation of magnetic moments with respect to the crystalline axes (including the optimization of the atomic coordinates). 
Figure \ref{fig:DFT} (a) shows $P_{c}$ as a function of the spin-rotation angle $\phi$: 
\color{black} according to what derived in  Eq. \ref{eqs:Pangle}, 
\color{black}
we observe a clear behavior, \color{black} 
 $P_{c}\propto\sin 2\phi$. 
As summarized in Tab.\ref{table:ptot}, 
the purely electronic contribution via SOC at fixed atomic structure, $P_{\rm elec}$=13$\mu$C/m$^{2}$, is strongly enhanced (up to 62$\mu$C/m$^{2}$) when  internal  atomic coordinates are optimized. 
This situation is similar to 
previously reported calculations on multiferroic TbMnO$_{3}$, where the purely electronic contribution $P^{\rm elec}$=32$\mu C/m^{2}$ 
is enhanced up to $P^{\rm elec+ion}$=-467$\mu C/m^{2}$ by  ionic relaxation.\cite{vanderbilt.tbmno3}  
In CaFeO$_{2}$, the maximum value of the calculated polarization is 62$\mu$C/m$^{2}$, slightly larger than the corresponding value in BCGO ($P$=47$\mu$C/m$^{2}$). 
Furthermore, MgFeO$_{2}$ is predicted to show a much larger polarization, $P$=327$\mu$C/m$^{2}$, comparable with TbMnO$_{3}$. 
\color{black} 
As shown in Tab.\ref{table:ptot}, the polarization shows a rather weak $U$-dependence; 
\color{black}
 the $P$ variation is approximately $\pm 15\%$. 
\color{black}
%
%
%

%
%
%
\begin{table}[h!]
\vspace{-0.2cm}
\caption{
Deviation of Fe-O-Fe bond angle from square-like configuration and $P_{c}$ under A$_{110}$ magnetic ordering, calculated in the fixed non-polar atomic structure, $P_{\rm exp}$ ($\mu$C/m$^{2}$)  and with optimized polar structure,  $P_{\rm opt}$ ($\mu$C/m$^{2}$). 
Experimentally,  $P_{c}$ in A$_{110}$ \bcgo\ is about 100$\mu$C/m$^{2}$\cite{tokura.prl2010}, 
whereas polarization in  CaFeO$_{2}$ and MgFeO$_{2}$ has not been measured. 
\color{black}
$P_{\rm opt}$ with different $U$ values (default $U$=4eV) are also shown.  
Note that structural parameters were fully optimized for each $U$ value. 
\color{black}
 \label{table:ptot}
}
\begin{center}
\begin{tabular}{|c|c|c|c|c|}
\hline
				&BCGO	&SrFeO$_{2}$	&CaFeO$_{2}$&MgFeO$_{2}$\\
\hline
180-$\angle$FeOFe ($^{\circ}$) &	---&0	&20.8	& 34.7\\
\hline
$P_{\rm fix}$   		&10		&0	&13	&22\\
\hline
$P_{\rm opt}$	&46		&0	&62 	&327	 \\
\hline
$P_{\rm opt}$ ($U$=3eV)		&49		&0	&58 		&284	 \\
$P_{\rm opt}$ ($U$=5eV)		&78		&0	&71		&378	 \\
\hline
\end{tabular}
\end{center}
%
%
\vspace{-0.5cm}
\end{table}
The strong increase of $P$ in MgFeO$_{2}$ is attributed to the larger O$_{4}$ tetrahedral distortion with respect to the centrosymmetric 
flat layer structure. 
In fact, we also performed  calculations on BeFeO$_{2}$, showing even more distorted tetrahedrons. 
Whereas we estimated a significantly large polarization, $P$$\sim$2000$\mu C/m^{2}$,  the chemical stability of the crystal structure with such a small ionic radius ion is questionable.  

\color{black}
It is shown in Tab.\ref{table:ptot} that the ionic contribution, $P_{\rm ion}$=$P_{\rm opt}$-$P_{\rm fix}$,  plays a crucial role in total $P$ 
\color{black}
and mostly comes from oxygen ionic displacements. 
%
When comparing our result for $P$ in iron-based oxides with our prototype, BCGO, we note that, 
from the structural point of view, 
\color{black} 
 FeO$_{4}$ tetrahedra are corner-shared, whereas 
 CoO$_{4}$ tetrahedra are intercalated by GeO$_{4}$ tetrahedra in BCGO. 
This leads to a cooperative effect between neighboring FeO$_{4}$ tetrahedra, so as to enhance the ionic displacement of O ion shared by two FeO$_{4}$ tetrahedra.

Ionic displacements driven under A$_{110}$ magnetic order in CaFeO$_{2}$ are  characterized by two phonon modes, $\Gamma_{1}$ and $\Gamma_{3}$. 
While the $\Gamma_{1}$ phonon mode doesn't change the original space group, the $\Gamma_{3}$ mode reduces the space group
from $P$-$42_{1}m1'$ to polar $Cmm21'$. 
Under the $\Gamma_{3}$ phonon mode,  MgFeO$_{2}$ exhibits 
 Mg  and O ionic displacements: both ions are shifted along the $z$ direction, 
by 0.0006 \AA\ 
and by 0.0011 \AA, respectively 
(displacements characterized by $A1(1)$  and $A'_{2}(a)$). 
%
\color{black}
It is therefore evident, once more, that there is a strong interplay among magnetism, ferroelectricity and lattice distortions. 
\color{black}
When one rotates the Fe spins from A$_{100}$ order to A$_{110}$ order, 
the  ions are  displaced so as to reduce SOC-induced stress in the structure. 
As previously shown in Tab.\ref{table:mae}, the in-plane MAE, 
$\Delta E= E(110; P\ne0) - E(100; P=0)$, is 0.00 and +0.60 meV/Fe for CaFeO$_{2}$  and  MgFeO$_{2}$, respectively. 
This implies that a sizable magnetic field is necessary to induce the polarization in MgFeO$_{2}$. 
However, the energy difference \color{black} related to in-plane MAE is reduced after  ionic relaxation: 
 $\Delta E^{\rm relax}$ goes to 0.00 meV/Fe.   
Therefore, owing to  ionic displacements occurring to reduce the SOC-induced  strain, 
the in-plane MAE becomes negligible,  guaranteeing an easy control of $P$ via applied magnetic field. 
\color{black}

\color{black}
Finally, we focus on polarization as induced by an applied field $H_{110}$. 
Starting from the A$_{1-10}$ AFM spin arrangement, we artificially rotated the spins by an angle $\phi'$ and evaluated polarization. 
Figure \ref{fig:DFT} (b) shows the results, confirming the prediction of Landau theory,  $P_{c}\propto \cos2(\phi')+\rm const$.
\color{black} 

\section{SOC effect}
\begin{figure}[ht]
\vspace{0cm}
{
\includegraphics[width=0.99\columnwidth, angle=0]{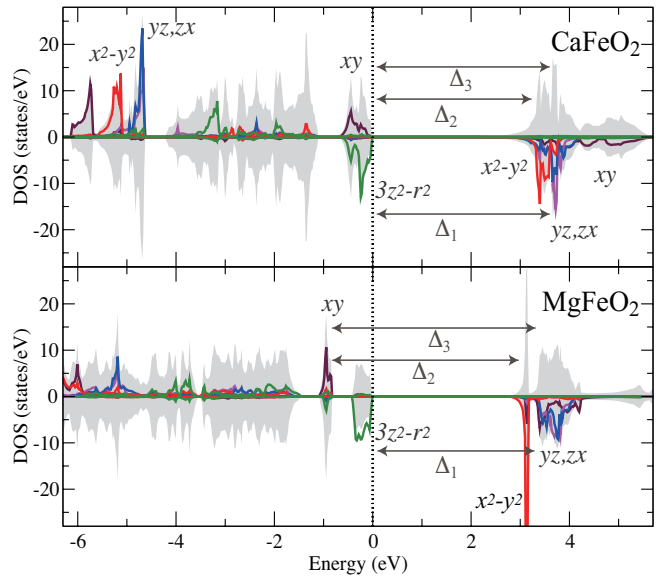}
}
\caption{\label{fig:dos} 
Density of states (DOS) around the Fermi energy ($E$=0) of CaFeO$_{2}$ and MgFeO$_{2}$. 
The $d$-orbital projected DOS 
and SOC mixing \color{black} parameters \color{black} between occupied and unoccupied states are labelled.  
}
\vspace{-0.3cm}
\end{figure}
%
%
\color{black}
The microscopic origin of the asymmetric $p$-$d$ hybridization  in CaFeO$_{2}$ can be explained via the same approach used in \bcgo, 
based on a tight-binding model considering a  cluster Hamiltonian of FeO$_{4}$ tetrahedron, 
$H =H_{d} +H_{p} +H_{pd} +H_{\rm SOC}$.\cite{yamauchi.bcgo} 
The $pd$ hybridization term, 
$H_{pd} = \sum_{\alpha, \beta, l}{V_{pd} (p^{\dagger}_{l, \beta}d_{\alpha} + h.c.)}$, 
is dependent on the oxygen ionic coordination and the Fe-$d$ occupied orbital state; 
the latter is tuned by Fe spin rotation via $H_{\rm SOC}$ term. 
The effect of   $H_{\rm SOC}$  term on the $d$ orbital states can be explained by treating SOC as perturbation. 
\color{black}

In the flat tetrahedral FeO$_{4}$ coordination (Fe is at 2$a$ site with $-4..$ symmetry), Fe-$d$ electrons occupy 
$3z^{2}-r^{2}$, $x^{2}-y^{2}$, degenerate $yz$/$zx$, and $xy$ orbital states in order, 
 consistently with the order of the nonbonding ($i.e.$ spatially avoiding O ligands) character.\cite{footnote0} 
Comparing the DOS of CaFeO$_{2}$ and MgFeO$_{2}$, 
the electronic structures are very similar,   
although the crystal field splitting between $3z^{2}-r^{2}$ and $x^{2}-y^{2}$ (=$\Delta_{1}$) is weakened 
in the more tetragonally distorted MgFeO$_{2}$.

\begin{figure}[ht]
\vspace{0cm}
{
\includegraphics[width=0.99\columnwidth, angle=0]{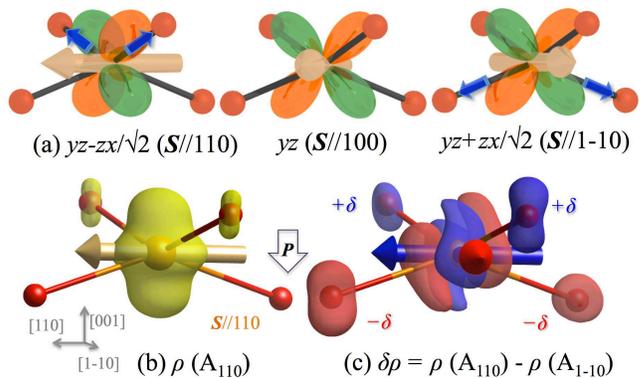}
}
\caption{\label{fig:diffchg} 
(a) Schematic picture of unoccupied Fe-$d$ orbital states mixed with occupied states under SOC. 
The orbital shape changes by rotating Fe spin.  
(b) Charge density isosurface of the highest occupied band (i.e.  Fe-$d$ 3$z^{2}$-$r^{2}$ orbital state) in MgFeO$_{2}$ under A$_{110}$ magnetic order. 
The spin direction is shown by an orange arrow. 
(c) Change in the charge density ($\delta \rho$) by rotating the Fe spin from [110] to [1-10] direction (positive shown in blue color; negative, in red). 
The direction of polarization caused by the charge difference ($\pm\delta$) is shown by an open arrow. 
For the sake of clarity in the pictures, SOC is artificially enhanced by factor of 10. 
}
\vspace{-0.3cm}
\end{figure}

The spin orbit coupling (SOC) term is described as $H_{\rm SOC}=\lambda\,\sum_{\alpha,\alpha'} \langle \alpha\vert\,L\cdot S\,\vert\alpha'\rangle  \, d^\dagger_\alpha
d^{\phantom{\dagger}}_{\alpha'}$, where the matrix elements can be expressed as a function of the polar and azimuthal
angles ($\theta,\phi$) defining a local reference for the spin-quantization axis.\cite{takayama.prb1976} 
\color{black}
The spin-orbit interaction mixes the occupied and unoccupied $d$-levels, so that the degeneracy of $yz$ and $zx$ states is removed: 
in parallel, an ``asymmetric hybridization'' of the latter orbitals with oxygens occurs, so that a local dipole moment along the $c$ direction arises. 
\color{black} 
Large SOC mixing is expected 
at $\Delta_{1}$ between $\left| 3z^{2}-r^{2}  \tiny{ \downarrow } \right>$ and 
\color{black}
 $\left| yz, zx  \tiny{  \downarrow } \right>$, 
\color{black}
$\Delta_{2}$ between $\left| xy \tiny{ \uparrow } \right>$ and $\left| x^{2}-y^{2}  \tiny{  \downarrow } \right>$ and 
$\Delta_{3}$ between $\left| xy \tiny{ \uparrow } \right>$ and $\left| yz, zx  \tiny{  \downarrow } \right>$, as shown in Fig.\ref{fig:dos}. 
Among them, $\Delta_{1}$ and $\Delta_{3}$ are more relevant to  polarization. 
Under this SOC-related mixing, the occupied states are modified by small contribution from the unoccupied states; \\
\begin{eqnarray}
\delta\left| 3z^{2}-r^{2} \tiny{ \downarrow } \right> &=  -\frac{\sqrt{3}i\lambda}{2\Delta_{1}}\left( \cos\phi  \left| yz\tiny{ \downarrow } \right>  -   \sin\phi  \left| zx\tiny{ \downarrow } \right> \right) \label{eq:4} \\
\delta\left| xy \tiny{ \uparrow } \right> &=  -\frac{\lambda}{2\Delta_{3}}\left( \cos\phi  \left| yz\tiny{ \downarrow } \right>  +   \sin\phi  \left| zx\tiny{ \downarrow } \right> \right),  \label{eq:5}
\end{eqnarray}
being $\theta$ set as 90$^{\circ}$. 
Comparing the coefficients in above Equations \ref{eq:4}-\ref{eq:5} 
and considering $\Delta_{1}$ and $\Delta_{3}$ to have a similar magnitude (see Fig.\ref{fig:dos}), we observe that the 
 SOC mixing involving $\Delta_{1}$ contributes to the change in orbital occupancy from the change in spin azimuth angle $\phi$ 
 by an amount which is about 3 times larger than the SOC mixing involving $\Delta_{3}$. 
 %
Therefore, the energetically highest occupied state $\left| 3z^{2}-r^{2} \tiny{ \downarrow } \right>$ is mixed with 
unoccupied $\left( \cos\phi  \left| yz\tiny{ \downarrow } \right>  +   \sin\phi  \left| zx\tiny{ \downarrow } \right> \right)$ orbital states, which modifies the shape according to the spin rotation (see Fig.\ref{fig:diffchg} (a)). 
This gives rise to the asymmetric $pd$ hybridization, 
enhancing the bonding character with upper oxygen states (when $\bf S$//[110]) or with lower oxygen states (when $\bf S$//[1-10]). 
\color{black}
It in turn moves the gravity center of Fe-$d$ charge;  
Fig. \ref{fig:diffchg} (c) shows 
the change in the DFT charge density of the occupied state by spin rotation calculation, 
which indeed is a visible proof of the asymmetric $pd$ hybridization mechanism. 
\color{black}
In this way, a local dipole, $p_c\propto \sin2\phi$,
 develops  along $c$, as   predicted from the functional form, 
$\bm{P}\propto\sum_{ij}(\bm{S}_{i}\,\cdot\,\bm e_{j}')^{2}\,\bm e_{j}'$. 
\cite{arima.jpsj2007,jia.nagaosa.prb2007,tokura.prl2010} 
%
\color{black} 
The latter also predicts $P_{c}$ to reach its maximum at $\phi=\pm 45 \degree$; i.e. when the direction of the Fe spin lies in the plane containing 
either the two upper of two lower oxygen bonds (cfr  Fig. \ref{fig:diffchg} (a)). 
%
We conclude by saying that, \color{black}  
although  CaFeO$_{2}$ and MgFeO$_{2}$ show a  similar electronic structure, 
the large enhancement of ME effect in  MgFeO$_{2}$ is due to the more distorted structure with respect to the centrosymmetric structure: 
as the CoO$_{4}$ coordination goes closer to a regular tetrahedron, 
the $p$-$d$ hybridization ($V_{pd}$) between Fe-$d_{yz, zx}$ and O-$p$ states becomes stronger. 
\color{black}
 
\section{Conclusions} On the grounds of  the known microscopic mechanism underlying peculiar magnetoelectricity observed in  Ba$_2$CoGe$_2$O$_7$, 
we predict much stronger magnetoelectric effects to appear in iron-based oxides, such as CaFeO$_{2}$, where a large polarization - magnetically controllable - is estimated.
In addition to essential ingredients, such as spin-dependent $p-d$ hybridization and spin-orbit coupling, here a central role is played by the peculiar 
 geometry, featuring ``flattened" FeO$_4$ tetrahedrons and non-centrosymmetric point group. The latter conditions are optimized in MgFeO$_2$, where our materials-design approach leads to magnified magnetoelectric effects, with a giant polarization tuned by magnetic fields.
 
\acknowledgments
The research leading to these results has received funding from 
 a Grant-in-Aid for Young Scientists (B)
from Japan Society for the Promotion of Science (JSPS) under Contract No. 24740235 
and 
from JST, CREST ``Creation of Innovative Functions of Intelligent Materials on the Basis of the Element Strategy''. 
The computation in this work has been done using the facilities of the
Supercomputer Center, Institute for Solid State Physics, University of Tokyo. 

\begin{thebibliography}{200}
\bibitem{footnote1}
\color{black}
In the Landau analysis, we also checked the WFM possibility, but we didn't find any polynomial terms of magnetic order parameters (such as $A_{i}F_{j}$ combination) in free energy. Therefore the spin-canting is not realized in this system and does not contribute to the ME effect. 
\color{black}
\bibitem{footnote0}
In this study, we set $x$, $y$, $z$ axis parallel to $a$, $b$, $c$ axis, respectively, as neglecting the small tilting of the O$_{4}$ tetrahedron in the $ab$ plane.
\end{thebibliography}
\end{document}